\documentclass[a4paper,fleqn,usenatbib]{mnras}

\usepackage{natbib}
\usepackage{multicol}

\usepackage{graphicx}	
\usepackage{amsmath}	
\usepackage{amssymb}	
\usepackage{subcaption}
\usepackage{float}
\usepackage{multirow}
\usepackage{bm}
\usepackage{mathabx}

\def\be{\begin{equation}}
\def\ee{\end{equation}}

\newcommand{\xobs}{{\bf x}_{\mbox{\scriptsize obs}}}
\newcommand{\xaux}{{\bf x}_{\mbox{\scriptsize aux}}}

\def\bea{\begin{eqnarray}}
\def\eea{\end{eqnarray}}

\usepackage{graphicx}
\usepackage{setspace}
\graphicspath{ {./images/} }
\PassOptionsToPackage{dvipsnames}{xcolor}
\RequirePackage{xcolor}

\usepackage{array}

\newcolumntype{M}[1]{>{\centering\arraybackslash}m{#1}}
\newcolumntype{N}{@{}m{0pt}@{}}

\DeclareGraphicsExtensions{.pdf,.png,.jpg,.eps}
\graphicspath{{figures/}}

\def\reff@jnl#1{{\rm#1\/}}

\def\aj{\reff@jnl{AJ}}                  
\def\araa{\reff@jnl{ARA\&A}}            
\def\apj{\reff@jnl{ApJ}}                        
\def\apjl{\reff@jnl{ApJ}}               
\def\apjs{\reff@jnl{ApJS}}              
\def\apss{\reff@jnl{Ap\&SS}}            
\def\aap{\reff@jnl{A\&A}}               
\def\aapr{\reff@jnl{A\&A~Rev.}}         
\def\aaps{\reff@jnl{A\&AS}}             
\def\baas{\reff@jnl{BAAS}}              
\def\jrasc{\reff@jnl{JRASC}}            
\def\memras{\reff@jnl{MmRAS}}           
\def\mnras{\reff@jnl{MNRAS}}            
\def\physrep{\reff@jnl{Phys.Rep.}}
\def\pra{\reff@jnl{Phys.Rev.A}}         
\def\prb{\reff@jnl{Phys.Rev.B}}         
\def\prc{\reff@jnl{Phys.Rev.C}}         
\def\prd{\reff@jnl{Phys.Rev.D}}         
\def\prl{\reff@jnl{Phys.Rev.Lett}}      
\def\pasp{\reff@jnl{PASP}}              
\def\pasj{\reff@jnl{PASJ}}              
\def\skytel{\reff@jnl{S\&T}}            
\def\solphys{\reff@jnl{Solar~Phys.}}    
\def\sovast{\reff@jnl{Soviet~Ast.}}     
\def\ssr{\reff@jnl{Space~Sci.Rev.}}     
\def\nat{\reff@jnl{Nature}}             

\newcommand{\beq}{\begin{equation}}
\newcommand{\eeq}{\end{equation}}
\newcommand{\beqa}{\begin{eqnarray}}
\newcommand{\eeqa}{\end{eqnarray}}

\usepackage{orcidlink}

\title[Fundamental Plane  ]{Reinterpreting Fundamental Plane Correlations with Machine Learning} 

\author[C.~Schafer et~al.]{
Chad Schafer$^{1,2,3}$\thanks{\tt cschafer@cmu.edu}, 
Sukhdeep Singh$^{2,3}$\orcidlink{https://orcid.org/0000-0001-6289-9208},
Yesukhei Jagvaral$^{2,3}$\orcidlink{https://orcid.org/0000-0001-7068-7037} 
\\
$^{1}$Department of Statistics \& Data Science, Carnegie Mellon University, Pittsburgh, PA 15213, USA \\
   $^{2}$McWilliams Center for Cosmology, Department of Physics, Carnegie Mellon University, Pittsburgh, PA 15213, USA \\
   $^{3}$NSF AI Planning Institute, Carnegie Mellon University, Pittsburgh, PA 15213, USA
}

\date{Accepted XXX. Received YYY; in original form ZZZ}

\pubyear{2023}

\begin{document}
\label{firstpage}
\pagerange{\pageref{firstpage}--\pageref{lastpage}}
\maketitle

\begin{abstract}

This work explores the relationships between galaxy sizes and related observable galaxy
properties in a large volume cosmological hydrodynamical simulation.
The objectives of this work are to both develop a better understanding of the correlations
between galaxy properties and the influence of environment on galaxy physics in order
to build an improved model for the galaxy sizes, building off of the  {\it fundamental plane}.
With an accurate intrinsic galaxy
size predictor, the residuals in the observed galaxy sizes can potentially be used for multiple cosmological applications, including making measurements of galaxy velocities in spectroscopic samples, estimating the rate of cosmic expansion, and
constraining the uncertainties in the photometric redshifts of galaxies. 
Using projection pursuit regression, the model accurately predicts intrinsic galaxy sizes and have residuals which have
limited correlation with galaxy properties. The model decreases the spatial correlation of galaxy size residuals by a factor of $\sim$ 5 at small scales compared to the baseline correlation when the mean size is used as a predictor.

\end{abstract}


\section{Introduction}

The difference between the intrinsic and observed (or inferred) size
of a galaxy is influenced by 
several physical processes, including gravitational lensing \citep{Bertin2006},
peculiar galaxy velocities \citep{Strauss1995}, doppler magnification \citep{Bonvin2017} and 
cosmic expansion \citep{Blakeslee2002}. With a sufficiently
accurate predictor of intrinsic
galaxy sizes, it is possible to construct estimators to study these effects using the size
residuals, i.e., the difference between observed and predicted intrisic
size. For example, the
anisotropies (the dipole) in the galaxy size cross correlations will be sensitive to the galaxy
velocities; the cross correlations of galaxy size with foreground galaxies are sensitive to
weak gravitational lensing caused by the foreground galaxies (galaxy-galaxy lensing cross
correlations); and the relation between the galaxy size and their redshifts can be used to
test the redshift-distance relation and hence models of cosmological expansion. Note that
these estimators use the size information differently and hence different measurements can
be carried out independently with only weak correlations/contamination between different
effects.

Such measurements also hold promise
to constrain the uncertainties in the
photometric redshifts of galaxies by exploiting the dependence of inferred galaxy size on
the estimated distance to the galaxy. The ratio of galaxy-galaxy lensing cross correlations
using the galaxy size residuals and galaxy shear is sensitive to the uncertainties in the
galaxy redshift estimates, i.e.
\begin{equation}
\frac{P_{g \lambda}}{P_{g \gamma}} - 1 \:\:\:\propto\:\:\: \delta \log D(z_{\mbox{\tiny source}})
\end{equation}
where $P$ is the cross power spectra (or correlation function), $g$ refers to the foreground lens galaxy,
$\lambda$
is the estimated size residual, $\gamma$ is the galaxy shear, and $\delta \log D(z_{\mbox{\tiny source}})$
is the error in the estimated
distance to the source galaxy (the galaxy for which we measure $\lambda$ and $\gamma$) due to uncertainties in
the photometric redshifts.
This estimation is similar to the consistency tests that have been done between galaxy shear (spin-2) and CMB convergence maps (spin-0), e.g. \cite{Singh2017cmb}, and estimators developed for such studies can be directly applied for comparing lensing measurements using galaxy shear and galaxy size (spin-0).   
Further, unlike the case of CMB lensing, since the size and shear are measured on the same
set of galaxies, the ratio is independent of the galaxy-matter power spectrum which is the primary
observable in the galaxy-lensing cross correlations, i.e. constraints on redshift will be almost
independent of the cosmological information. 
Independence from galaxy-matter power spectrum
implies that the measurement is also independent of the cosmic variance and will only depend on
the measurement noise and intrinsic scatter in the size and shear measurements. Since photometric
redshift uncertainties are one of the limiting systematics when analyzing data from photometric galaxy
surveys, including galaxy size estimates can potentially lead to significant improvements in
cosmological inferences, beyond a simple improvement in statistical errors.

An accurate and precise predictor of intrinsic
galaxy size minimizes the scatter in the size residuals, which is the primary source of
noise in cosmological measurements. One such size predictor is the fundamental plane (FP) of
galaxies \citep{fp-1,fp-2}. 
FP is the relation between the size, $R_0$, surface brightness, $I_0$ and velocity
dispersion, $\sigma_0$, of elliptical galaxies given by
\begin{equation}
\log R_0 = a \log \sigma_0 + b \log I_0 + c + \sum_{i=1}^{N_z} d_i z_i
\end{equation}
where the redshift, $z_i$, dependent terms were introduced in \citep{Joachimi2015} to account for the
redshift evolution of the plane \citep[see also discussion in][]{Singh-2021}. While studied extensively
in the literature in the context of galaxy physics, a careful study of the FP in context of cosmological
measurements has only recently gained traction \citep[e.g.][]{Joachimi2015, Saulder2019, Singh-2021} 
and the efficacy of the FP for cosmological analysis is not well established.

\cite{Singh-2021} performed a detailed study of the FP residuals and the galaxy
properties involved in the FP definition. The FP residuals were found to be strongly correlated
with the galaxy properties, e.g. the mean of the FP residuals increases with galaxy luminosity. These
correlations suggest that the scatter over the FP is not strictly random. Furthermore, the FP
residuals are correlated with the galaxy density field, an effect similar to the intrinsic alignments
of galaxy shapes. This effect can also be explained by the dependence of the galaxy properties
on their environment. Brighter and larger galaxies tend to reside in over-dense regions, though \cite{Singh-2021}
observed that these galaxies have lower surface brightness. Correlations of FP with these properties
explains the correlations of FP residuals with the galaxy density field.
For cosmological applications, it is important to understand these correlations of galaxy properties in order to improve the galaxy size predictors and avoid biases in the cosmological
inferences. The physical origins of these correlations are still not well understood and better understanding of these effects is important to improving models of galaxy physics. 

This work explores such correlations 
using state-of-the-art, large cosmological volume hydrodynamical simulations and performs a more detailed study to understand the correlations between galaxy sizes and several other galaxy properties. 
The use of a simulation model (IllustrisTNG, described
below in 
Section \ref{simulation}) for this purpose enables a
more thorough exploration of correlations with a
wider range of galaxy properties, measured with minimal
error. Of course, the ultimate objective is to use these
models with observed data, and a focus of this work is to develop novel methods
of analysis which will enable this by using the high-resolution
information available from the simulation model to guide
the fitted model.

Meeting the above objectives motivates the development of novel
analysis methodology for incorporating the rich structural information
obtained
from large simulation models, and this is also a focus of this work.
A fundamental question is the
following: Suppose that some feature of the galaxy or its
environment (e.g., a measure of 3D density) is known to be
useful in predicting intrinsic galaxy size, but that such
information is only available in a simulation model. Is there
a way to exploit the relationship between 3D density
and other {\bf observable} galaxy properties to better predict
galaxy size? One may believe that sophisticated supervised
learning methods should be capable of discovering the optimal model
for the relationship between observable properties and
galaxy size, but the complexity of this model may make it difficult
to ascertain, and difficult to interpret.
We place emphasis here on an approach that balances interpretability
and predictive power. The simulation model provides a useful
framework around which models can be built that are not of
excessive complexity, but achieve strong prediction performance.

The remainder of this paper is organized as follows: Section
\ref{data} describes the simulation model utilized, and the
galaxy and environment features derived from it.
Section \ref{methods} presents the statistical tools behind
the model and its assessment.
Section \ref{results} describes the primary model fit in this
work.
Section \ref{discussion} discusses the results and its
implications for future exploration.

\section{Data}\label{data}
\subsection{The cosmological simulation}\label{simulation}

IllustrisTNG \citep{tng-bimodal,pillepich2018illustristng, Springel2017illustristng, Naiman2018illustristng, Marinacci2017illustristng,tng-publicdata} comprises cosmological hydrodynamical simulations that were run with the moving-mesh code Arepo \citep{arepo}. 
The TNG100 simulation at $z=0$ was chosen for this study since the simulation exhibits
color bimodality that agrees with SDSS data for intermediate mass galaxies \citep{tng_2color}, as well as consistent correlations with other galaxy properties. Additionally, TNG100 provides a good balance between high resolution and a large cosmological volume. 

The box of 75 Mpc/h $\sim$ 100 Mpc has $2 \times 1820^3$ resolution elements with a gravitational softening length of 0.7 kpc/h 
for dark matter and star particles.  
The mass of dark matter and star particles are $7.46\times 10^6  M_\odot$ and $1.39\times 10^6  M_\odot$, respectively.  Additionally, the simulation incorporates various physical process for galactic evolution:  radiative gas cooling and heating; star formation in the ISM; stellar evolution with metal enrichment from supernovae; stellar, AGN and blackhole feedback; formation and accretion of supermassive blackholes \citep{tng-methods,tng-agn}.

The dark matter halos were identified using the friends-of-friends (FoF) 
algorithm \citep{fof}, and then the subhalos  
were identified using the SUBFIND algorithm \citep{subfind}. 
 We  employ a minimum stellar mass cut of $ \log_{10}(M_*/M_\odot) =9 $ , roughly corresponding to $10^3$ star particles \citep{ Tenneti_2016,Du_2020}.

\subsection{Galaxy and Environment Properties}\label{props}

This section characterizes the source of
galaxy properties used in the predictive models.
Some standard quantities utilized, such as \textit{size (half-mass radius)} and \textit{star formation rate} arrive directly from the simulation catalog; for more information on these, we refer the interested reader to the simulation model website\footnote{{\tt https://www.tng-project.org/}}. The \textit{velocity dispersion} of each individual galaxies was calculated using the velocities of all star particles in a galaxy.
 
\vspace{.1in}
\noindent \textit{Density Measures.} In the models below, both
2D and 3D galaxy density information is utilized.
To calculate 2D density, galaxy counts are tabulated
on a 1000 by 1000 grid, and then smoothed using a
Gaussian kernel with a scale of 0.5 Mpc/h. This density,
evaluated at the galaxy positions, is stored as
{\tt delta\_smooth\_R}.
Similarly, for 3D density,
galaxy counts are tabulated on a
$750 \times 750 \times 750$ grid, and then smoothed using a Gaussian
kernel with scales of 0.5, 1.0, 2.0, and 5.0 Mpc/h. This generates
measures of density, at varying scales, for the environment local to
each galaxy.

 \vspace{.1in}
 \noindent \textit{Galaxy Morphological Classification.} Galaxy morphology is
 characterized using the probabilistic dynamical model of \cite{gal_decomp}.
The model makes two physically motivated assumptions.
First, it is assumed that the angular momentum of disc stars is approximately aligned with the total angular momentum of the galaxy, while the angular momentum of bulge stars angular is randomly aligned.
Second, it is assumed that the orbits of disc stars are approximately circular,
while the orbits of bulge stars orbits are elongated or circular.

 In order to quantitatively model the aforementioned assumptions, define the following:
\begin{itemize}
  \item  $j_\text{r} \equiv \frac{j_\text{star}} {j_\text{circ}(r)}$, where $\bm{j}_\text{star}$ is the angular momentum of a single star particle and $j_\text{star}$ is its magnitude;   $j_\text{circ}(r) = r\, v_\text{circ}(r) = r \, \sqrt{\frac{GM(r)}{r}}$   is the expected angular momentum for a circular orbit at the same position as that star,
where $M(r)$ is the total mass (across all types of particles -- stars, gas, dark matter) contained within that radius.   

  \item $\cos\alpha$ is the cosine of the angle between the angular momentum vector of the star particle and the total angular momentum of the galaxy. 
  \end{itemize}

Next consider the following model for the distribution of star particles:
\begin{eqnarray}
   p_\text{star}(j_\text{r},\cos\alpha) & \hspace{-.1in}\equiv& \nonumber \\ 
   & &
   \hspace{-.9in}
   (1-f^\text{disc}) \, p_\text{bulge} (j_\text{r},\cos\alpha) + f^\text{disc} \, p_\text{disc}  (j_\text{r} , \cos\alpha).
\end{eqnarray}
Here, $p_\text{bulge}$ and $p_\text{disc}$ are the densities (both normalized to integrate to 1) reflecting the probability that a star at a given point in this 2D space belongs to the bulge or to the disc. More details and further investigations of the model can be found in \cite{gal_decomp}. Finally, {\tt mc\_disk}, or the galaxy disk fraction, is calculated by adding up the mass of
all of the star particles that were classified as disks and dividing by the total mass.

\section{Methods}\label{methods}

As stated above, this work is focused not only on developing improved models for predicting galaxy
size from measurable quantities, but also on providing better understanding of the relationships
between these properties.
Hence, a methodological focus of this work is to utilize approaches that balance modelling accuracy
with scientific
interpretability. This section will discuss the use of projection pursuit regression as an
alternative to neural networks and other machine learning approaches. Ultimately, the residuals from
these fits must be analyzed to determine if there are remaining correlations with intrinsic 
galaxy properties, hence this section will also discuss methods for such approaches.

The {\em projection pursuit regression (PPR) model} \citep{Friedman:1980tu} is characterized as follows. The response variable $Y$ is
modelled as a additive combination of $m$ different nonlinearly-transformed projections of
the predictor vector ${\bf x}$:
\begin{equation}
   Y_i = \sum_{j=1}^m \beta_j f_j\!\!\left(\balpha_j^T {\bf x}_i\right) + \epsilon_i.
\end{equation}
The $\epsilon$ are
assumed to be mean zero, uncorrelated {\em irreducible errors}, i.e.
scatter around the model fit. Here, the $\beta$, the
$\balpha_j$,
and the $f_j$ are {\it estimated} from the available training sample.

The $\balpha_j$ represent the $m$ different projections of the original predictors ${\bf x}_i$ that are utilized by the model.
This approach avoids the
{\it curse of dimensionality} by only considering an additive
combination of what could be viewed as {\em designed features}
$f_j\!\!\left(\balpha_j^T {\bf x}_i\right)$ for
$j=1,2,\ldots,m$.
The model has the flexibility to learn the
linear combinations of the predictors $\balpha_j$, in tandem
with the nonlinear transformation $f_j$, which are the most
useful for predicting the response. 
The $f_j$ will typically be estimated
via standard nonparametric regression approaches such as with a smoothing
spline \citep{Reinsch1967SmoothingBS}.
Such approaches are well-suited to one-dimensional regression
problems such as this since they can flexibly fit to a wide
range of relationships (here, between $\balpha_j^T {\bf x}_i$
and the response). Such fits are smooth, but allow the
data to dictate the shape of the fit, i.e.,
no parametric form is assumed.

It is instructive to contrast the projection pursuit model with a 
{\em fully-connected single layer neural network} model
\begin{equation}
   Y_i = \sum_{j=1}^m \beta_j \phi\!\left({\bf w}_j^T {\bf x}_i\right) + \epsilon_i,
\end{equation}
wherein the
user fixes an {\em activation function} $\phi$, a simple nonlinear
transformation which is applied
to each (of typically many) linear combinations of the predictor
vector. The parameters learned from the data are solely the values of the weights ${\bf w}_j$ applied in these linear combinations.
Projection pursuit is able to use smaller $m$ by
exploiting the flexibility in the tailored, nonlinear transformation
$f_j$ that is applied to each. This leads to improvements in
interpretability.

The model is fit using a two-level iterative approach.
An outer loop consists of running over $k=1,2,\ldots,m$.
For fixed $k$, the residuals from the fit on the other
$m-1$ components is calculated:
\begin{equation}
    r_i = Y_i - \left[\sum_{j \neq k} \widehat \beta_j \widehat f_j\!\!\left(\widehat\balpha_j^T {\bf x}_i\right)\right]
\end{equation}
and then $\beta_k$, $f_k$, and $\balpha_k$ are found such
that
\begin{equation}
r_i \approx \beta_k f_k\!\!\left(\balpha_k^T {\bf x}_i\right).
\end{equation}
This relationship between the $r_i$ and $(\beta_k, f_k, \balpha_k)$ uses
an inner loop which alternates between estimating
$\beta_k f_k$ and $\balpha_k$. Heuristically, at this step
the goal is to determine how to best fit to the portion of
response that is unexplained by the other $m-1$ terms in the
model. With each update to a $(\beta_k, f_k, \balpha_k)$,
the other components are eventually reconsidered as the outer
loop is repeated until convergence is reached.
This procedure referred to as {\it backfitting}
\citep{Breiman1985}.

{\bf Comment: Implementation.} In this work, models are fit using the function {\tt ProjectionPursuitRegressor} found in of Scikit-learn.
\citep{scikit-learn}.
{\it Smoothing splines} \citep{wahba1990spline} are used in each
one-dimensional nonparametric fit $f_j$, of
degree either two or three. (Initial
models are fit using cubic splines, but
in the final model degree will be chosen
as part of the procedure described below.)
The seemingly-redundant parameters $\beta_j$ are included in the model reflecting the
custom of {\tt ProjectionPursuitRegressor}
and other software.
This was not a part of the original formulation of \cite{Friedman:1980tu}
but allows for extra generality in simultaneous
fitting of multiple response vectors using the same collection of $m$ ridge functions.

\subsection{Preliminary Models}

\begin{figure*}
  \center
  \includegraphics[width=\textwidth]{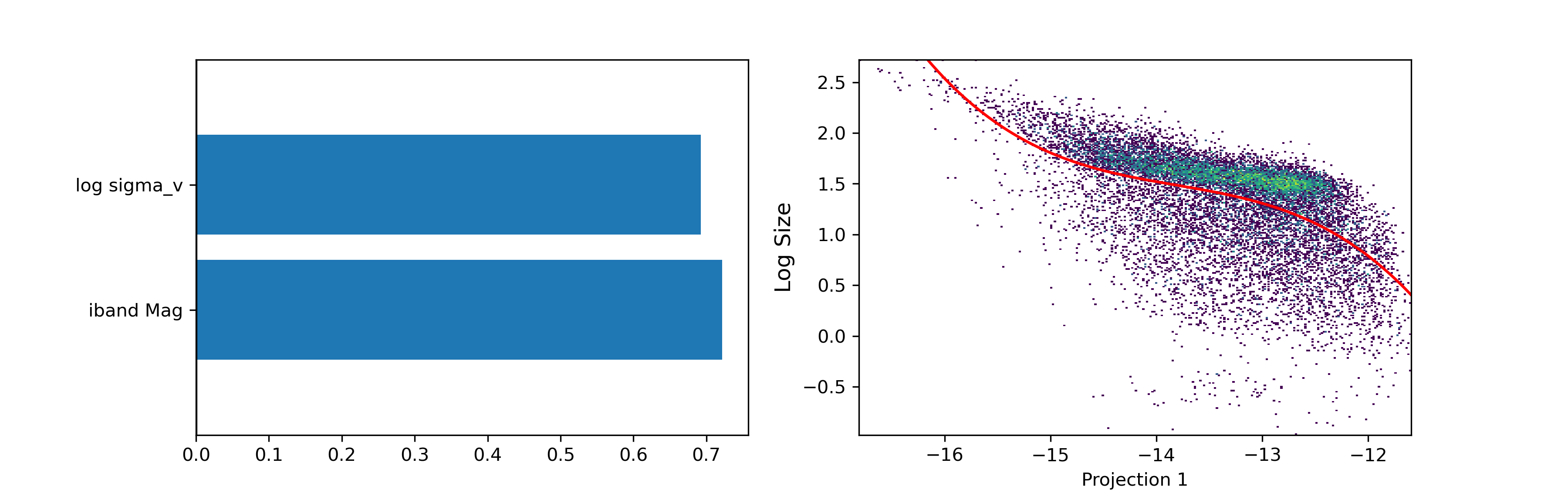}
  \caption{Illustration of the results from the first, simple fit. Here, $m=1$, and there
  are only two predictors, log velocity dispersion and i-band magnitude. The left figure shows the weight placed on each of these two predictors, while the right shows the non-linear function applied to this linear combination.}
    \label{OrigFunPlaneFit}
\end{figure*}

As an initial demonstration, a model is fit with log radius as the response, and log velocity
dispersion and i-band magnitude as predictors, to mimic the classic fundamental plane model.
Here, $m=1$, a special case of projection pursuit called {\it single index regression}.
Figure \ref{OrigFunPlaneFit} illustrates the results. The left panel shows the weight placed
on each of the two predictors to form the first projection, i.e., $\widehat \alpha_{11} = 0.69$
and $\widehat \alpha_{12}= 0.72$. The horizontal axis in the right panel shows the value of
this projection for all observations in the training set. The estimated form for $\beta_1 f_1$
is shown as the solid curve on the panel. This is fit using a cubic spline. The quality of this
simple fit is clearly poor (with RMSE on a test set of 0.411), with deficiencies partly due to the range of different galaxy
types being fit.

\begin{figure*}
  \center
  \includegraphics[width=\textwidth]{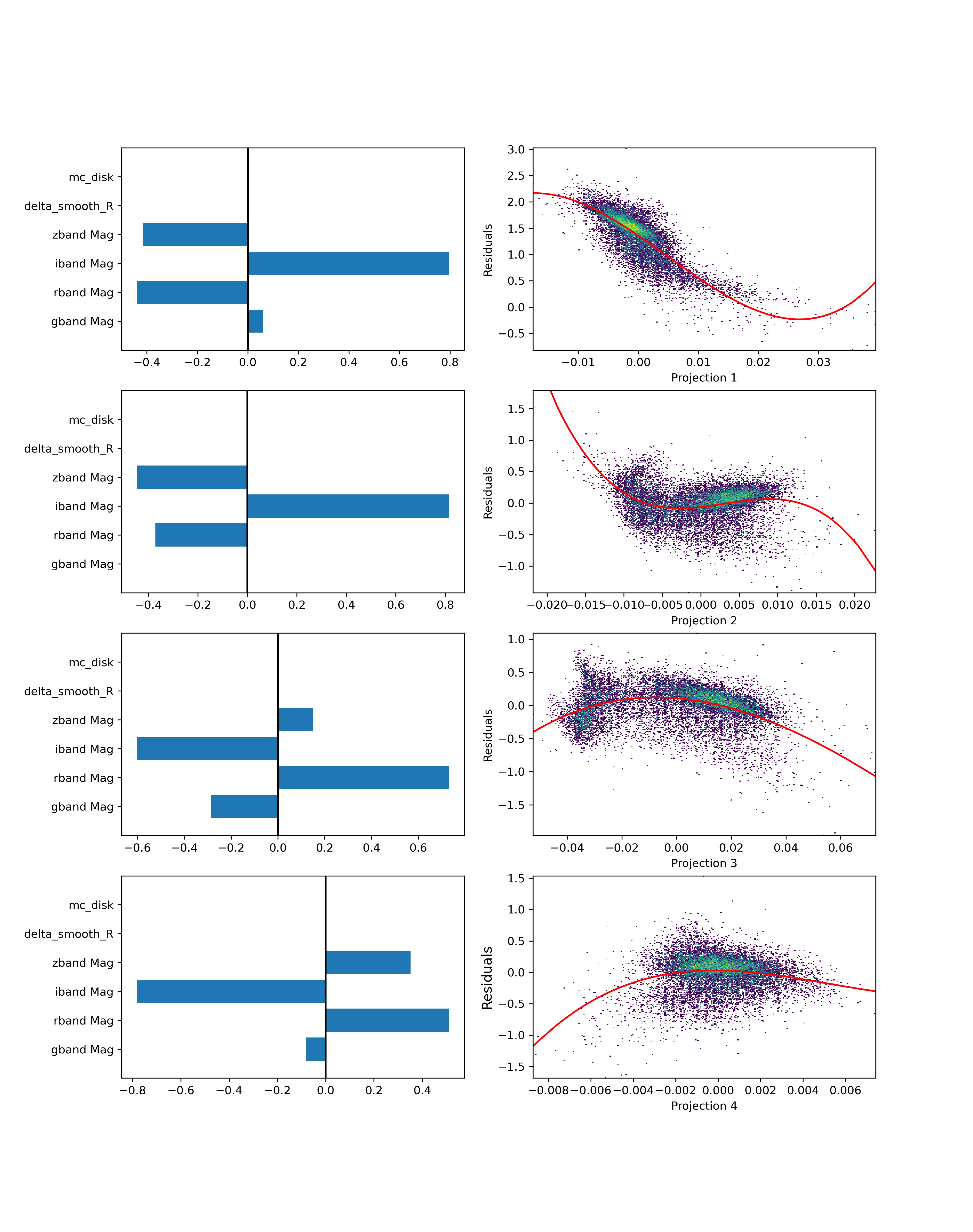}
  \caption{Illustration of the results from the second fit. Here, there are six predictors (all based on photometry), and $m=4$. The vertical axes of right
  plots is labelled ``Residuals'' because
  the figure shows the fit to what remains
  after the other $m-1=3$ components is
  subtracted off.}
    \label{Model2fit}
\end{figure*}

A primary motivation of this work is to build models for intrinsic
galaxy size that can be used in cases where only photometric observations
are available. Hence, for comparison, next is fit a model that includes
the $griz$ magnitudes as the features, along
with {\tt mc\_disk} and {\tt delta\_smooth\_R}, described above in Section \ref{props}.
Each feature is individually shifted and
scaled to have mean zero and standard deviation one prior to the fit. When $m=1$, the RMSE on a
test set is 0.290, but this improves to 0.257 when $m=4$. The
results are shown in Figure \ref{Model2fit}. It is notable
that the model appears to place little weight on {\tt mc\_disk} and {\tt delta\_smooth\_R}, but, in fact, excluding these two predictors increases the test set RMSE to 0.272.
 
{\bf Comment: Splitting the Data.} Throughout this work,
when data are divided into training, test, or other sets for the purposes
of model fitting and validation, splits are done {\em by pixels} formed
in a $5 \times 5 \times 5$ grid that covers
the full simulation box of 75 Mpc/h. This is
done to mitigate issues that could
result from galaxies in close proximity sharing important physical
information, and hence inappropriately influencing the quality of the model fit.

\subsection{Residual Analysis}

The RMSE values reported for each model only partially reveal important information regarding the quality of the fit, because minimizing prediction errors is not the primary objective
of this work.
To serve the cosmological motivations, the ideal model would leave
no remaining relationship between the residuals from the fit
and any properties of the galaxy and its environment. In other words, the model would
predict intrinsic size, and the difference between the measured size and the fit size
would encode useful information regarding gravitational lensing,
peculiar galaxy velocities, and so forth. To this end, study of the property of the
model residuals is crucial.

\begin{figure*}
  \center
  \includegraphics[width=\textwidth]{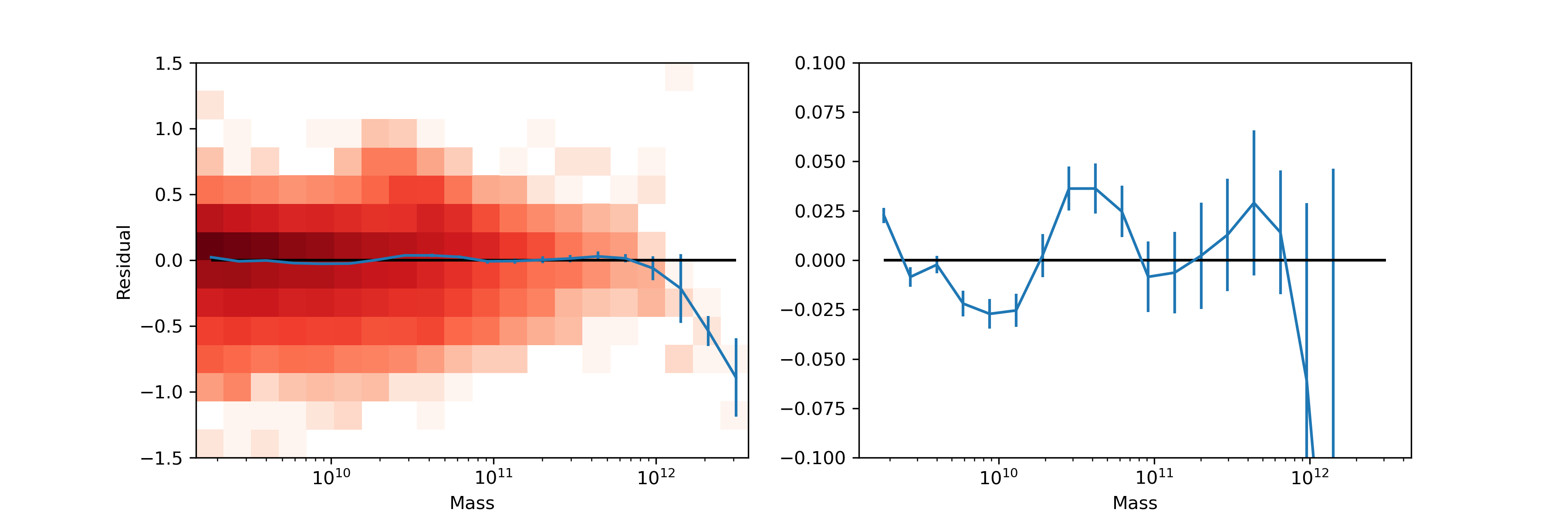}
  \caption{Evolution of the residuals with galaxy mass. The mean residual in bins is shown
  in the blue curve. Error bars are constructed using a jackknife procedure.}
    \label{residplot1}
\end{figure*}

One step in this direction is to plot residuals versus various galaxy properties, and
look for patterns and/or trends. Figure \ref{residplot1} shows the result of comparing
galaxy mass with the residuals from the model fit above to photometry-based properties. The right panel shows 
the clear evolution in
residuals with galaxy mass, where the blue curve shows mean residuals in each of
20 bins. Error bars shown are calculated on each of these means using a jackknife procedure, described below.
Similar comparisons can be made with other features, including those which are included in the model. This is an important step in
revealing deficiencies in the model fit.

\begin{figure*}
  \center
  \includegraphics[width=0.6\textwidth]{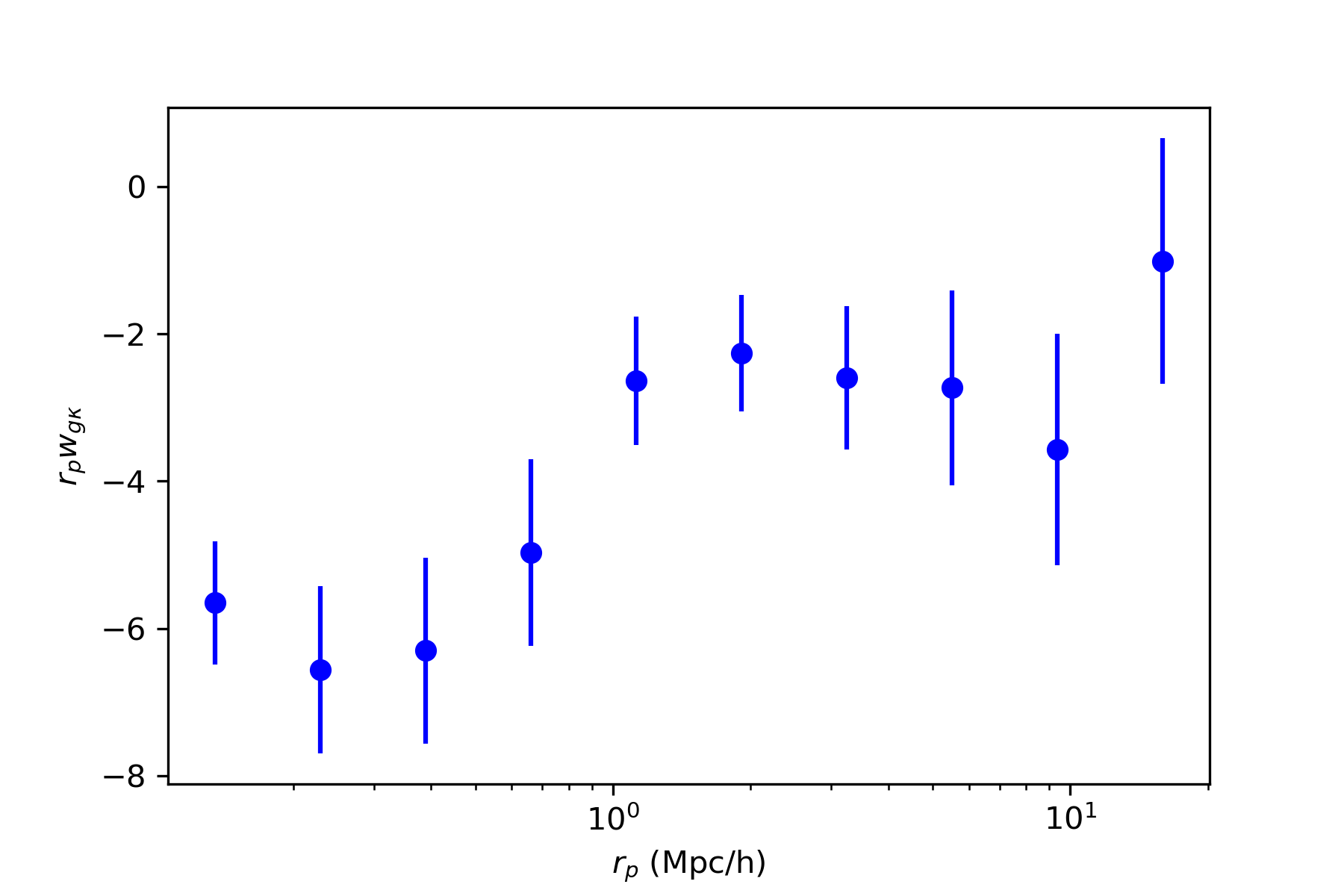}
  \caption{Measurements of correlations between the galaxy size residuals and the galaxy density field. This effect is similar to the intrinsic alignments effect for galaxy shear. It arises because size residuals are a non-linear combination of galaxy properties which are correlated with the galaxy environment.}
    \label{residplot2}
\end{figure*}

Figure \ref{residplot2} assesses the degree
of spatial correlation in the residuals. Such spatial correlations have been analyzed in the previous cosmological studies using fundamental plane of galaxies \citep{Joachimi2015, Singh-2021}. These correlations exist because  the galaxy formation and evolution involves complex physical processes that depend not only on the galaxy itself but also its environment. These contaminate the estimators that are used in cosmological measurements of interest using size residuals and it is desirable to null them before cosmological analysis.
Again, in an ideal model there would be no
remaining spatial correlation in the residuals from
the model fit. However, in figure \ref{residplot2} we see strong correlations between the size residuals and the surrounding galaxy density field. Such signal is not totally unexpected as the size residuals are a non-linear combination of galaxy properties which are correlated with the environment \citep[see ][ for detailed explanation and analysis]{Singh-2021}.
This correlation of galaxy sizes with the local density field is very similar to the intrinsic alignments effect for galaxy shear. Unfortunately, this implies that the current size estimators cannot be used to perform the cosmological measurements using auto-correlations, but they are suitable for cross correlations, similar to galaxy-shear cross correlations. 

{\bf Comment: Jackknife Errors.}
The classic {\it jackknife} \citep{EfronStein1981}
approach to calculating errors
on estimators $\widehat \theta$
consists of repeatedly re-calculating the
estimate, each time leaving out one observation from
the sample.
If the estimate when observation $i$ is removed is
denoted as $\widehat \theta_{(-i)}$, then
\begin{equation*}
  \widehat {\rm Var}_{\mbox{\tiny jk}}(\widehat \theta) =
  \left(\frac{n-1}{n}\right)
  \sum_{i=1}^n \!\left(\widehat \theta_{(-i)} - 
  \overline{\widehat \theta}\right)^2
\end{equation*}
can be shown to be a reliable estimator of the true
variance of $\widehat \theta$, in the case where
the sample is drawn independent and identically
distributed from some population.
In this work, this assumption is clearly invalid
due to dependencies present from nearby galaxies.
Hence, the jackknife procedure is adapted to one
in which the 3D simulation box is divided into a
grid of $7^3$ pixels, with each pixel left out in one iteration of the procedure. This reduces the bias
that would result from taking the standard jackknife approach of leaving out one observation (galaxy) at a time.

\subsection{Kernel PCA-Based Enhancement of PPR}

As described above, the PPR model is built upon linear
combinations of the supplied collection of features chosen to be optimal
for predicting the targeted response variable. Hence, this model can be viewed as
a {\it supervised} companion to principal components analysis (PCA), wherein a new
representation of data vectors is constructed in an {\it unsupervised} manner, with a
goal of finding linear combinations with maximal variance. This is motivated by
the heuristic that
directions in the original space along which there is the greatest variability
are the projections that encode the most useful information.
Thus, standard PCA used in combination with projection pursuit regression would
be redundant, as nothing could be gained by considering a simple rotation of the
features in Euclidean space. There exists, however, a nonlinear extension of PCA,
called {\it Kernel PCA} \citep{Scholkopf1998NonlinearCA}
which provides a potentially useful enhancement to the space of
projections under consideration by the PPR model.

It is instructive to
first consider the math behind standard (linear) PCA.
For additional detail, see \cite{hastie2009elements}. Let ${\bf X}$ denote the $n$ by $p$ matrix whose rows
are the individual feature vectors. Assume that the variables have been mean-centered so that each column of ${\bf X}$ has sample mean zero, hence ${\bf X}'{\bf X}/n$
is the sample covariance matrix for these data. Then, the principal
components are found as the eigenvectors of ${\bf X}'{\bf X}/n$, or, equivalently, of
${\bf X}'{\bf X}$. Denote these eigenvectors as ${\bf v}_1, {\bf v}_2, \ldots, {\bf v}_p$
and the corresponding eigenvalues with $\lambda_i$. PCA can
be interpreted as creating a new coordinate system, or basis, within which a data
vector can be represented so that ${\bf s}_i = {\bf X} {\bf v}_i$ provides the positions of all
$n$ observations along the
$i^{th}$ axis in the new coordinate system.
It follows that
\begin{equation}
    {\bf X}'{\bf X} {\bf v}_i = \lambda_i {\bf v}_i
    \:\mbox{,}\:\:\:\:
    {\bf X} {\bf X}'{\bf X} {\bf v}_i = \lambda_i {\bf X} {\bf v}_i
    \:\:\:\:\mbox{and}\:\:\:\:
    {\bf X} {\bf X}'{\bf s}_i = \lambda_i  {\bf s}_i.
\end{equation}
Since $\|{\bf s}_i\|^2 = \lambda_i$,
the standardized versions $\mathsf{s}_i = {\bf s}_i/\sqrt{\lambda_i}$ will be orthonormal
and hence are the eigenvectors
of 
the
{\it Gram matrix} ${\bf X}{\bf X}'$. The conclusion
is that the position in the new coordinate system can be found directly from the eigenvectors of the Gram matrix.

In Kernel PCA, this form of the Gram matrix is generalized such that
the $(i,j)$ entry is $K({\bf x}_i, {\bf x}_j)$, where $K$ is a user-chosen {\it kernel function} which measures similarity between vectors. Common choices for the kernel function
include the {\it radial basis function kernel}
\begin{equation}
    K({\bf x}, {\bf y}) = \exp\!\left(-\gamma \| {\bf x}-{\bf y}\|^2\right)
\end{equation}
and the {\it sigmoid kernel}
\begin{equation}\label{sigmoidkernel}
    K({\bf x}, {\bf y}) = \tanh\!\left(\gamma {\bf x}'{\bf y} + c\right).
\end{equation}
Both of these examples illustrate the important role of tuning parameters in the choice of
a kernel, e.g., through the specification of $\gamma$.

Kernel PCA also maps the observations into a
new space, with the hope that a useful lower-dimensional
representation will result.
Let $\bphi({\bf x})$ denote the position
of ${\bf x}$ in the new space defined by Kernel PCA.
The $i^{th}$ coordinate is found via the {\it Nystr{\"o}m extension} \citep{Nystrom},
\begin{equation}
    \phi_i({\bf x}) = \frac{1}{\sqrt{\lambda_i}} \sum_j  \mathsf{s}_{ij} K({\bf x}, {\bf x}_j).
\end{equation}
For ${\bf x}_k$ included in the training set,
$\phi_i({\bf x}_k) = \sqrt{\lambda_i}\:\mathsf{s}_{ik}$.

The connection with PPR is as follows: A linear combination of the predictors,
now using the Kernel PCA representation, is
\begin{equation}
    \balpha'\!\bphi({\bf x}) = 
   \sum_j  \left(\balpha'\!\mathsf{s}_{\cdot j} \right) K({\bf x}, {\bf x}_j),
   \label{lincomb}
\end{equation} 
where $\mathsf{s}_{\cdot j}$ holds $\mathsf{s}_{ij}$ for $i =1,2,\ldots,n$.
(In this expression, $\lambda_i$ are absorbed into the individual $\alpha_i$
without loss of generality.) The heuristic behind this is that varying
the tuning parameter $\gamma$ that
characterizes the kernel function
leads to a wide range of nonlinear
transformations of the predictor
vector. The model can achieve
a better fit if it has a larger class of {\it intelligently-chosen} directions to
search over. The vector $\bphi({\bf x})$ can be of dimension up to $n$, while the
original predictor was limited to $p$ dimensions. The choice of $\gamma$ allows for
great flexibility in the formation of this new representation.

An analogous idea is often employed with a standard
approach to classification,
{\em support vector machines (SVM)} \citep{Cortes1995SupportVectorN}.
The basic SVM approach searches for a linear
separator in the feature space to distinguish the two classes under consideration. Of course, a linear
separator in the original feature space
is rarely an adequate classifier. But by projecting
the features into a much higher-dimensional space, the
potential for finding a useful linear separator is
greatly enhanced. This is often referred to as the {\em kernel trick}.

\begin{figure*}
  \center
  \includegraphics[width=0.7\textwidth]{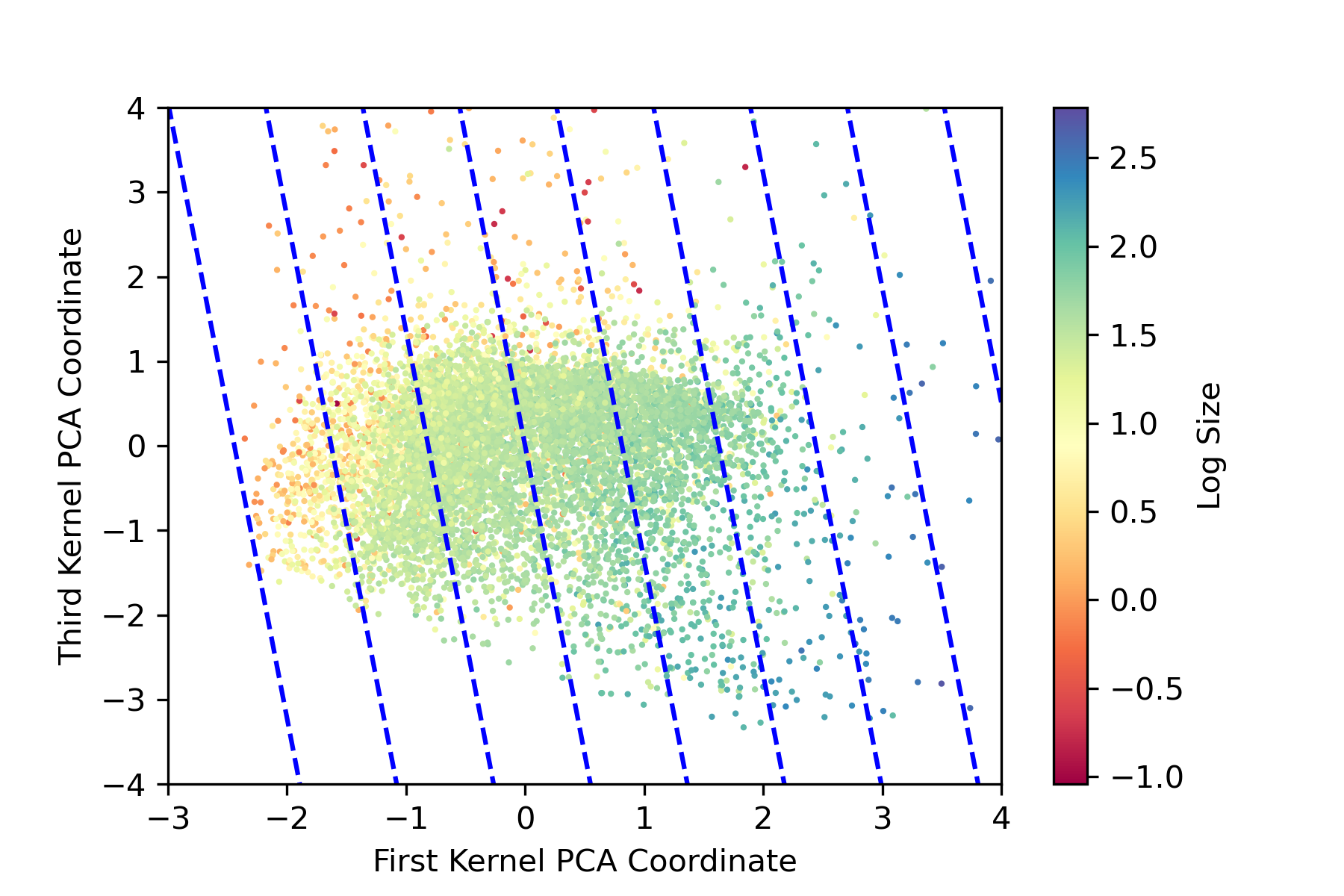}
  \caption{This figure illustrates how the inclusion of Kernel PCA coordinates enhances the fit. The position of each galaxy along the first and third Kernel PCA coordinates is shown, colored by the galaxy size. The dashed contours show lines of constant galaxy size, as fit by the PPR model.}
    \label{ProjDemo}
\end{figure*}

Figure \ref{ProjDemo} illustrates the potential. In this fit, Kernel PCA was used to create a nonlinear transformation of galaxy properties into a 
ten-dimensional space. Galaxy properties utilized were photomery-based, as in the previous fit, but now in the PPR model these features are supplemented with those derived from Kernel PCA. The sigmoid kernel was used. The figure shows how the PPR model is able
to exploit this new representation to find a direction in the new space along which the response evolves. The dashed lines show contours which the model is fitting to have constant log size. It is important to keep in mind
that this shows one such projection; in fact, $m=4$ in this model, so there are four such projections through the ten-dimensional space created by Kernel PCA. The RMSE on a test set is reduced to 0.240.

\subsubsection{Incorporation of Auxiliary Information}

In the application of interest, ${\bf x}$ will be decomposed into the pair ${\bf x} = [\xobs, \xaux]$. Here, $\xobs$ consists of the observable properties of the galaxy,
i.e., quantities that can be measured or adequately-estimated using solely photometry.
The variables in $\xaux$ will be additional properties of a galaxy that are not
observable, but are believed to encode information useful for predicting its size.
These include properties such as three-dimensional density information and the galaxy's
location in the central or satellite region of its cluster. This {\it auxiliary
information} is unobservable in photometric surveys, but will be available in a high-resolution simulation model such as the
Illustris model used in this study. The objective is here is to exploit this additional information to improve predictions of intrinsic galaxy size.

The approach developed here will build off the Kernel
PCA-enhanced PPR model described above. First, note
that in Equation \ref{lincomb}, the $\mathsf{s}_{\cdot j}$ are not dependent on the particular ${\bf x}$ for which the prediction is
sought. These $\mathsf{s}_{\cdot j}$ represent the
{\em directions} found in the new, Kernel-PCA
derived representation of the features. Since they
are not dependent on ${\bf x}$, these can be
learned from a training set that has full access
to both $\xobs$ and $\xaux$, e.g., the information
generated from a simulation model. The dependence
on ${\bf x}$ arises only in the kernel function
$K({\bf x},{\bf x}_j)$. The additional complication
of this approach comes from the need to
approximate $K({\bf x},{\bf x}_j)$ from
$K({\xobs},{\bf x}_j)$.

Hence, in the first step
the {\it auxiliary training set} is constructed from the simulation
model, i.e., for these $n_a$ galaxies both $\xobs$ and $\xaux$ are
available. From this set, a low-dimensional representation is learned
using the Kernel PCA approach described above. The tuning parameters of the chosen kernel function will become
tuning parameters for the final prediction model.
The result of this first step is a set of vectors (directions) $\mathsf{s}_{\cdot j}$
for $j=1,2,\ldots, n_a$.
Again, these directions can exploit
the rich information available in
the features in both $\xobs$ and
$\xaux$.

For the next step, recall from above that the position of any ${\bf x}$ in this new space
can be found as
\begin{equation}\label{newpos}
    \phi_i({\bf x}) = \frac{1}{\sqrt{\lambda_i}} \sum_j  \mathsf{s}_{ij}
    K({\bf x}, {\bf x}_j).
\end{equation}
The challenge at this point is that on the actual, observed data, only $\xobs$
is available. To account for this, the kernel function $K$ will be approximated
via
\begin{equation}
    \widehat K(\xobs, {\bf x}_j) \approx K([\xobs,\xaux], {\bf x}_j).
\end{equation}
Here, this approximation will be achieved using a neural network
learned from the auxiliary training set derived from the
simulation model.

The natural question at this point is the following: What is gained by the
incorporation of the auxiliary information? I.e., is it not possible to
simply model the galaxy size as a function of $\xobs$ directly through
a model? This is definitely possible, but this approach is exploiting
the additional structure available in the auxiliary information. The
auxiliary variables are demonstrably quite powerful source
for making these predictions. This information is passed on through
the vectors $\mathsf{s}_{\cdot j}$ which are learned from the auxiliary
training set.

A second evident question is as follows: Would it be better to fit
one or more models that learn the relationship between $\xobs$ and
$\xaux$, use these to impute the unavailable $\xaux$ vectors,
and then use these in a model trained on the auxiliary training
set? The approach advocated for here avoids the fitting of several
models, or one model with a vector-valued response, and instead
focuses directly on approximating a single, real-valued quantity
which encodes the important information, namely the kernel function
evaluated at relevant pairs.

In Section \ref{results} below, results are presented from fitting using this procedure.

\section{Models for Galaxy Size}\label{results}

This section will present the results from the fitting of a more
sophisticated model for intrinsic galaxy size. The approach will follow
what is outlined in Section \ref{methods}, with a mix of features
from simulation model and photometric sources, all used in an
effort to build an improved model for the size.
The features based on photometry are as above: The $griz$ magnitudes,
{\tt mc\_disk}, and {\tt delta\_smooth\_R}. (These
latter two quantities are described in Section \ref{props}.)
The auxiliary features extracted
from the simulation model are as follows: galaxy mass, velocity dispersion, star formation rate, 3D density measures, and central versus satellite classification of the galaxy's location
within its cluster.

\subsection{Model Pipeline Architecture}

\begin{figure*}
  \center
  \includegraphics[width=0.6\textwidth]{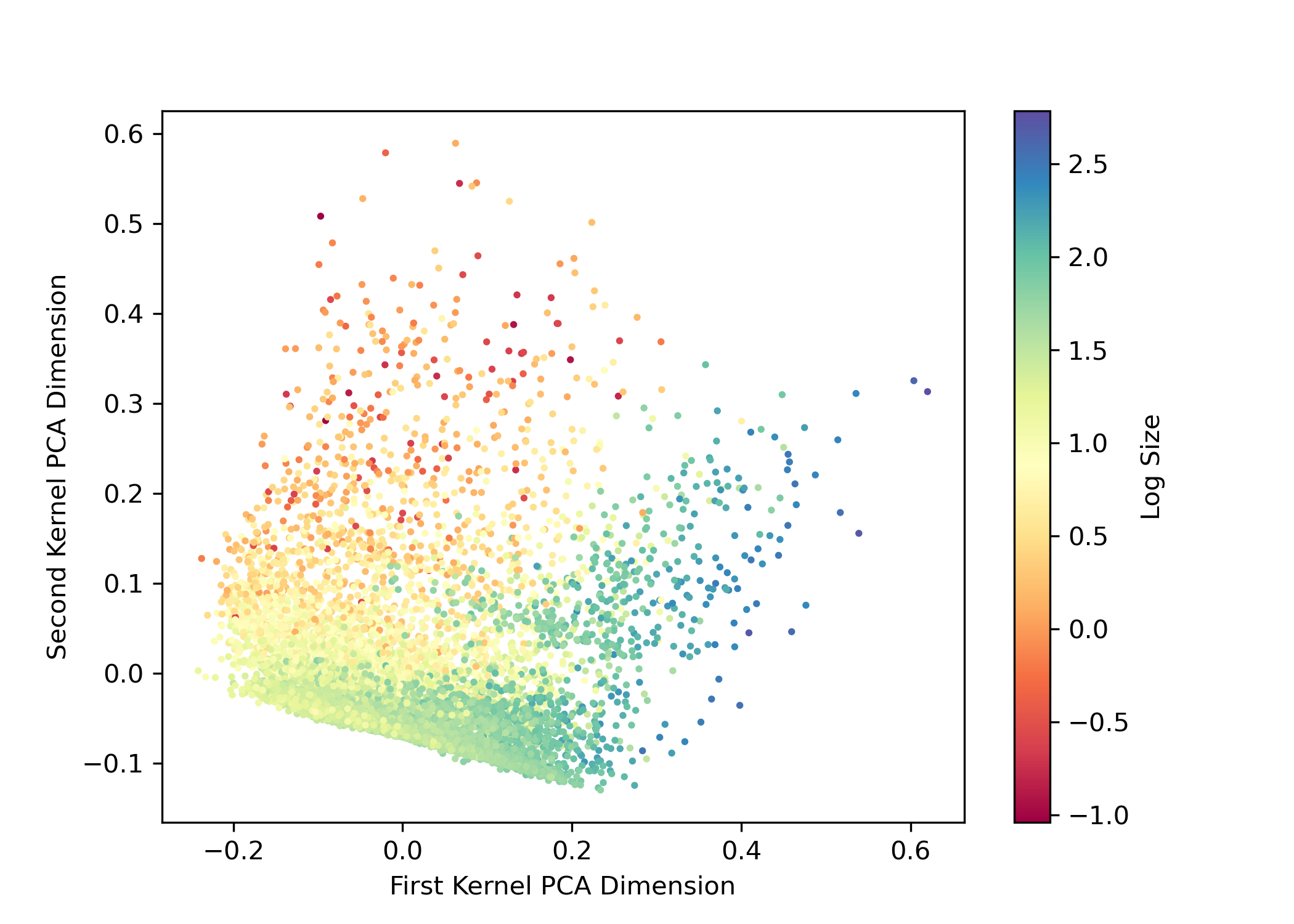}
  \caption{The first two dimensions created by the Kernel PCA
  transformation. Each dot represents one galaxy, and color reflects the log size. The evident relationship between position in this
  space and galaxy size suggests that Kernel PCA is picking up
  important physical information.}
    \label{KPCAresults}
\end{figure*}

For the purposes of this modelling pipeline, the data are divided into three sets. (As mentioned above, groups are formed by pixel.) First, {\bf Set 0} consists of those galaxies used to create the
Kernel PCA representation. Here, this is done using the sigmoid kernel
(Equation \ref{sigmoidkernel}) with $c=1$ and with $\gamma$ the first of the tuning parameters to be optimized. (The approach
to setting the values of the tuning parameters is described below.) Figure \ref{KPCAresults} depicts the
first two dimensions in this representation, showing
the important relationship with galaxy size. Ultimately, the number
of dimensions which are used in the model is another tuning
parameter.

\begin{figure*}
  \center
  \includegraphics[width=0.4\textwidth]{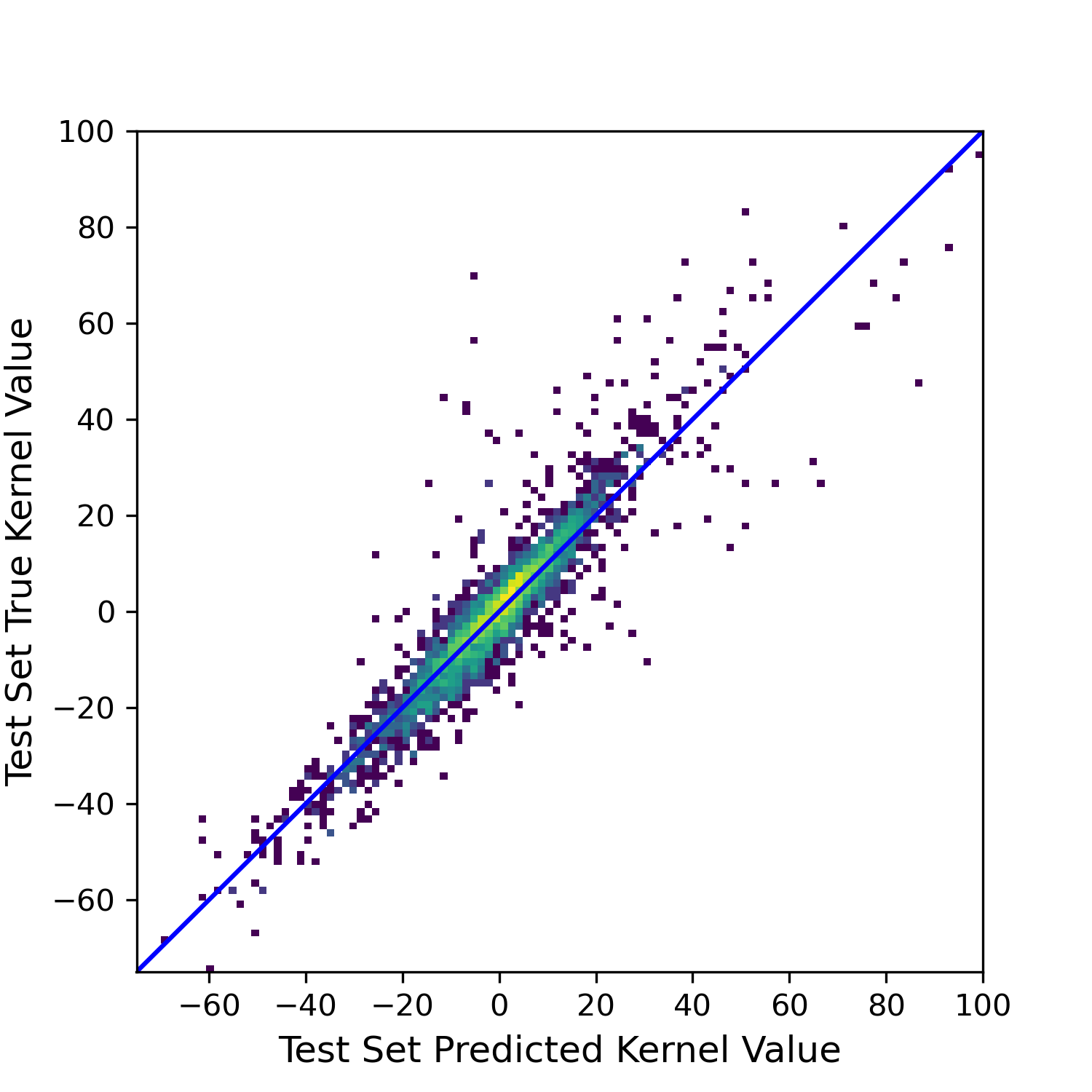}
  \caption{Comparison of the actual and predicted kernel values when
  using a neural network model.}
    \label{ANNresults}
\end{figure*}

In the next step, the observations in Set 0 are further divided into a training and test set for the purposes of predicting the kernel function when evaluated at a $(\xobs, {\bf x})$ pair. This model is fit using a
fully-connected, four-layer neural network, with 1000 nodes per layer. 
Learning is allowed to run for 200 epochs, with
learning rate fixed at 0.001.
The dropout rate (applied after each
layer) and the mini-batch size in the utilized ADAM
optimizer are additional tuning parameters.
Figure \ref{ANNresults} shows the performance of
this model on the test set in the final chosen model.

The role of the aforementioned model is to allow for the prediction
of the
value of $K(\xobs, {\bf x})$ for pairs where $\xobs$ is {\bf not}
in Set 0, but ${\bf x}$ is from a
galaxy which is in Set 0. To understand this
step, it is useful to consider an updated
version of Equation
\ref{newpos} above, as follows:
\begin{equation}\label{newpos}
    \widehat \phi_i(\xobs) = \frac{1}{\sqrt{\lambda_i}} \sum_j  \mathsf{s}_{ij}
    \widehat K(\xobs, {\bf x}_j).
\end{equation}
Here, $\widehat \phi_i(\xobs)$ is the position in the $i^{th}$ dimension of the Kernel PCA of the galaxy with
observed properties $\xobs$, when the approximated
Kernel is utilized. One can imagine that the Set 0 galaxies
comprise a collection of simulation model-derived
``reference points'' to which the
galaxies outside Set 0 are compared, albeit using an approximation
to the Kernel function. The values of $\widehat \phi_i(\xobs)$ is
calculated relative to these reference points.

At this stage, the information is available for fitting the projection pursuit regression model that relates galaxy size (on the log scale)
to the mix of observable and Kernel PCA-generated
features.
{\bf Set 1} is the training set used for this model,
while {\bf Set 2} is held out as a test set.
In this model, the number of Kernel PCA features,
the degree of the spline functions, and the number of ridge
functions are tuning parameters.

\subsection{Selection of Tuning Parameters}

A challenging aspect of fitting a model of this
complexity is the number of tuning parameters that
result. In this pipeline, some model components are
fixed to values that are deemed to be reasonable,
e.g., the use of 1000 units in each layer of the
neural network, and the choice of the sigmoid kernel.
Other tuning parameters are set via
randomization at the outset of the pipeline:
\begin{itemize}
    \item The value of $\gamma$ in the Kernel PCA
    procedure, with $\log_{10}(\gamma)$ chosen
    uniformly on the interval $(-5, -2)$.
    \item The dropout rate used in the neural
    network model, chosen uniformly between 0.1
    and 0.5. (The same dropout rate is used for
    all four layers.)
    \item The batch size used in the neural network
    fitting algorithm, set to 16, 32, 64, or 128.
    \item The number of Kernel PCA dimensions used
    in the PPR fit, set to 10, 15, or 20.
    \item The degree of the spline functions in the
    PPR fit, set to either 2 or 3.
\end{itemize}
The final tuning parameter is the number of ridge
functions used in the PPR model. With each of the
above five parameters fixed, this is varied from 2 to
16, with a cross-validation approach used to
choose its value.
Ultimately, the figure of merit used in choosing
the global set of tuning parameters is the minimal
MSE within this cross-validation procedure.
The values chosen by this procedure
are as follows: $\gamma$ equals $0.00535$,
dropout rate equals $0.22$, batch size equals 64,
there are ten retained KPCA dimensions, and quadratic splines are used in the PPR model.

{\bf Comments:} As an additional hedge against
overfitting, this cross-validation procedure
uses the {\it one-SE rule} 
 \citep{James2013}, wherein
the value of the associated tuning parameter is
set to the smallest such value which yields a figure of merit within one
standard error of the best performing choice.
The motivation behind this approach is that one should only
choose a model if there is convincing evidence
that the additional complexity is warranted.
Also, note that this procedure avoids using the test set (Set 2) in the selection of the tuning parameters,
which helps to preserve the role of the test set
as an ultimate tool for assessing the performance
of the model.


\subsection{Model Performance}

\begin{figure*}
  \center
  \includegraphics[width=\textwidth]{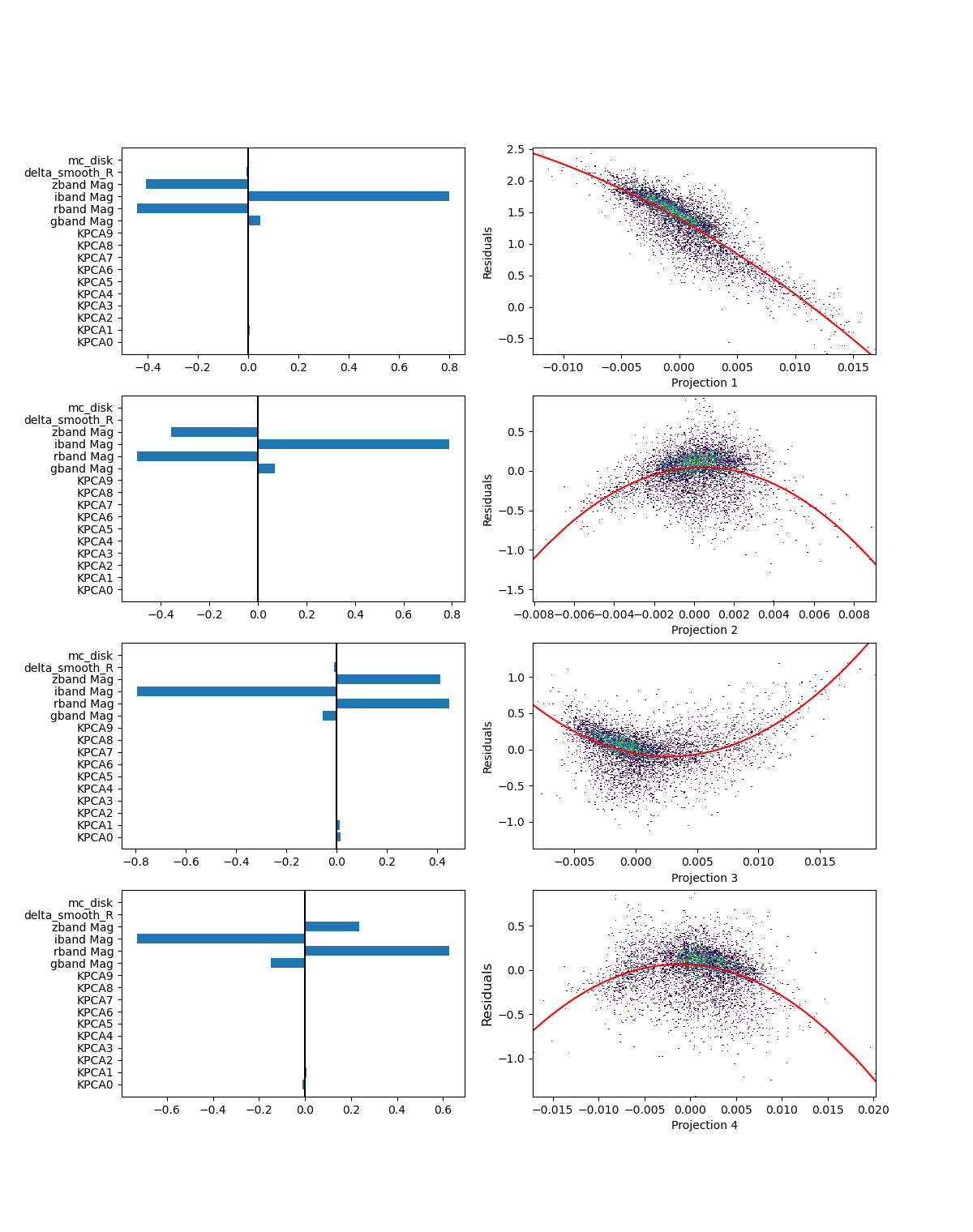}
  \caption{Illustration of the first four ridge functions fit in the final model, following the optimization of the tuning parameters. It is evident from the
  weights placed on the $griz$ magnitudes that
  these are the most important in predicting the
  size of the galaxy. The kernel-PCA derived
  factors are also influencing the fit, however,
  and help to reduce the RMSE error on a test set
  to 0.231.}
    \label{FinalModelFits0}
\end{figure*}

\begin{figure*}
  \center
  \includegraphics[width=\textwidth]{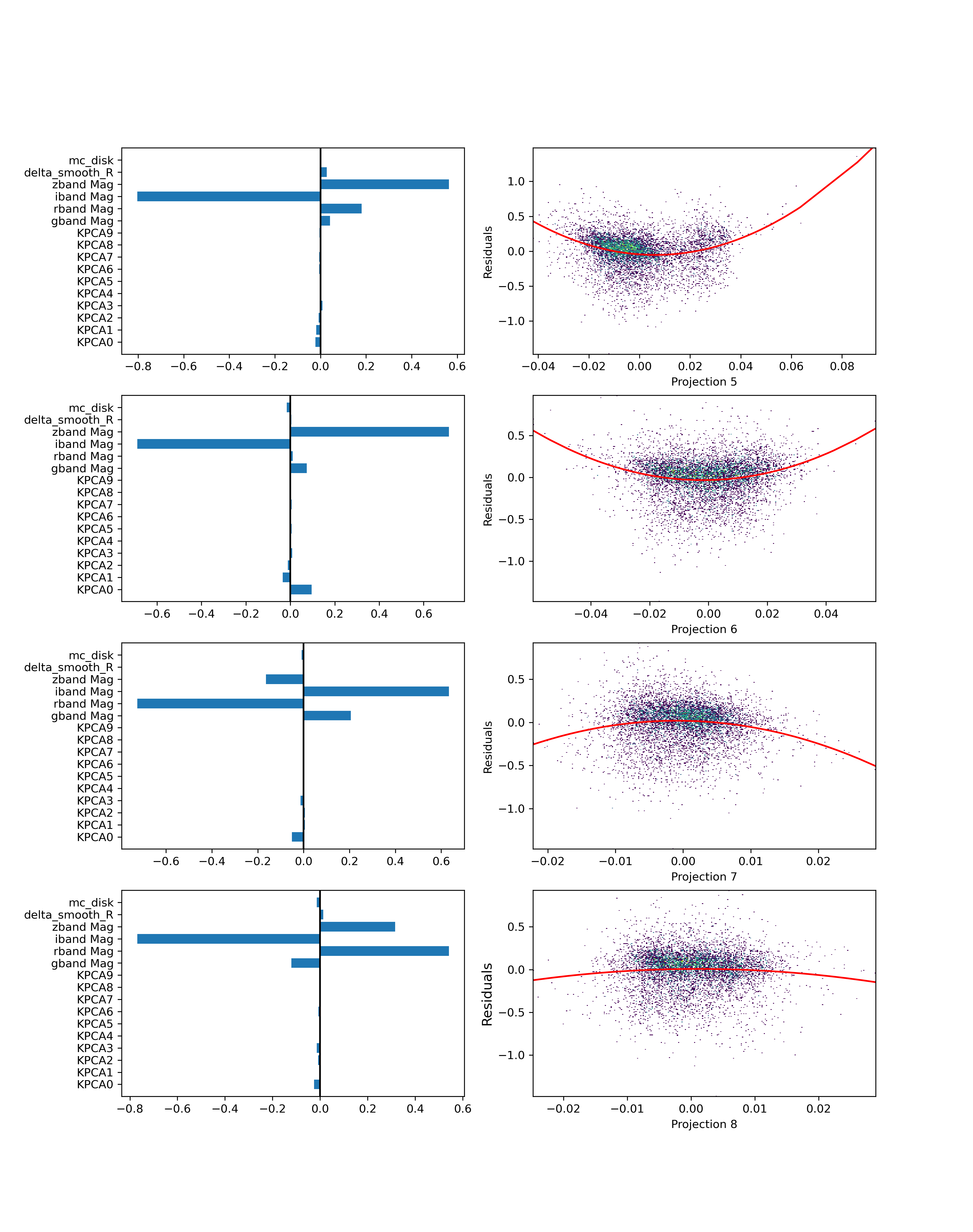}
  \caption{Illustration of the second four ridge functions fit in the final model, following the optimization of the tuning parameters. It is evident from the
  weights placed on the $griz$ magnitudes that
  these are the most important in predicting the
  size of the galaxy. The kernel-PCA derived
  factors are also influencing the fit, however,
  and help to reduce the RMSE error on a test set
  to 0.231.}
    \label{FinalModelFits1}
\end{figure*}

\begin{figure*}
  \center
  \includegraphics[width=0.5\textwidth]{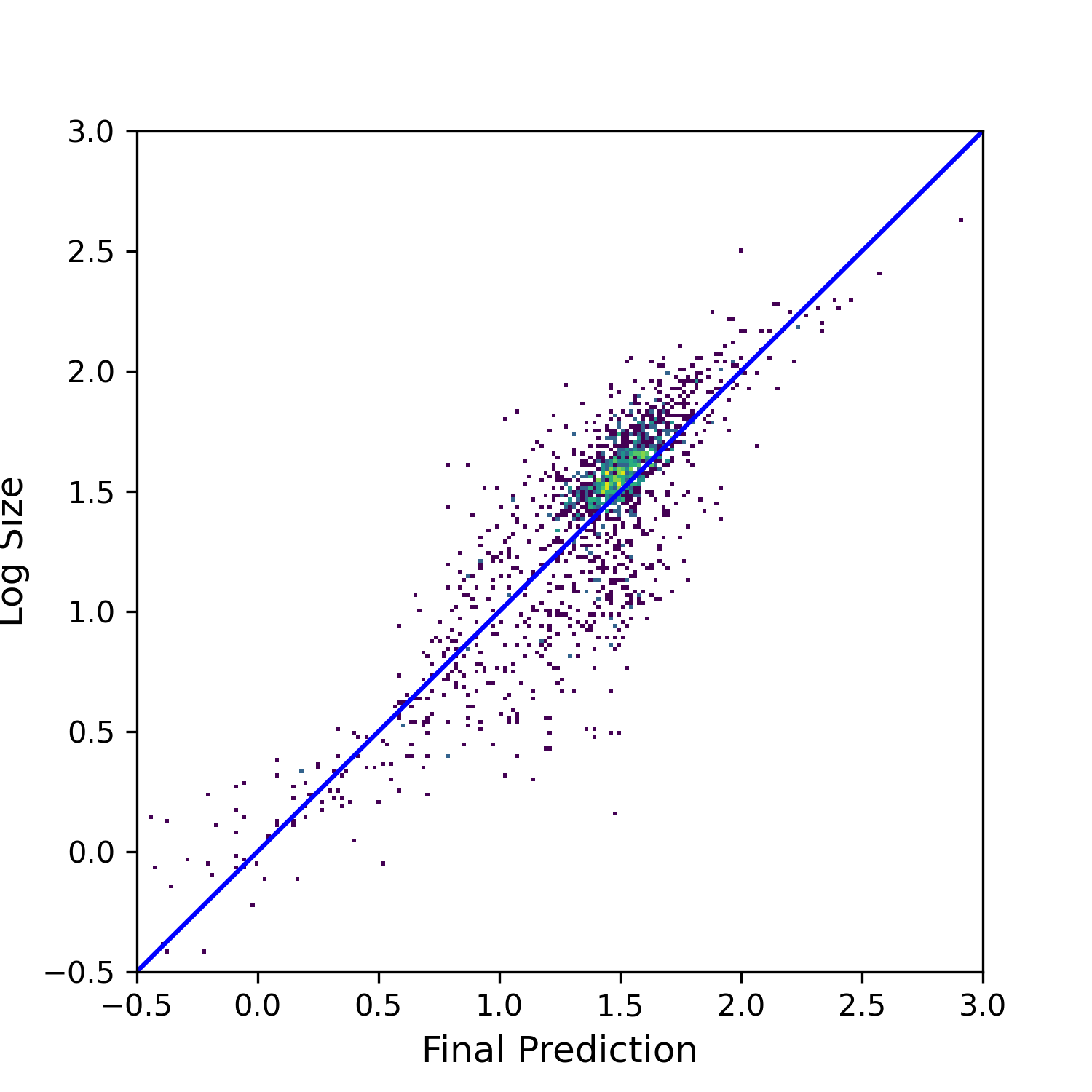}
  \caption{Illustration of the results from the full model fit. The true galaxy size on the test set
  is compared with the predictions from the model.
  The RMSE in the scatter around the depicted
  line of 
  agreement is 0.231.}
    \label{FinalModelTestSet}
\end{figure*}

Figures \ref{FinalModelFits0} and \ref{FinalModelFits1} show the eight ridge functions
fit in this model.
The results show that the contrasts in the
magnitudes (i.e., the colors) are clearly the
most important in predicting the response values. Each of the kernel PCA-derived
predictors receive small weight, but they
still play a crucial role in improving
the predictions, as evidenced by the
reduction in the RMSE on the test set
error to 0.231.
Figure \ref{FinalModelTestSet}
compares the model predictions with the true
galaxy size, for observations in test
test set, i.e., Set 2.

\begin{figure*}
  \center
  \includegraphics[width=0.90\textwidth]{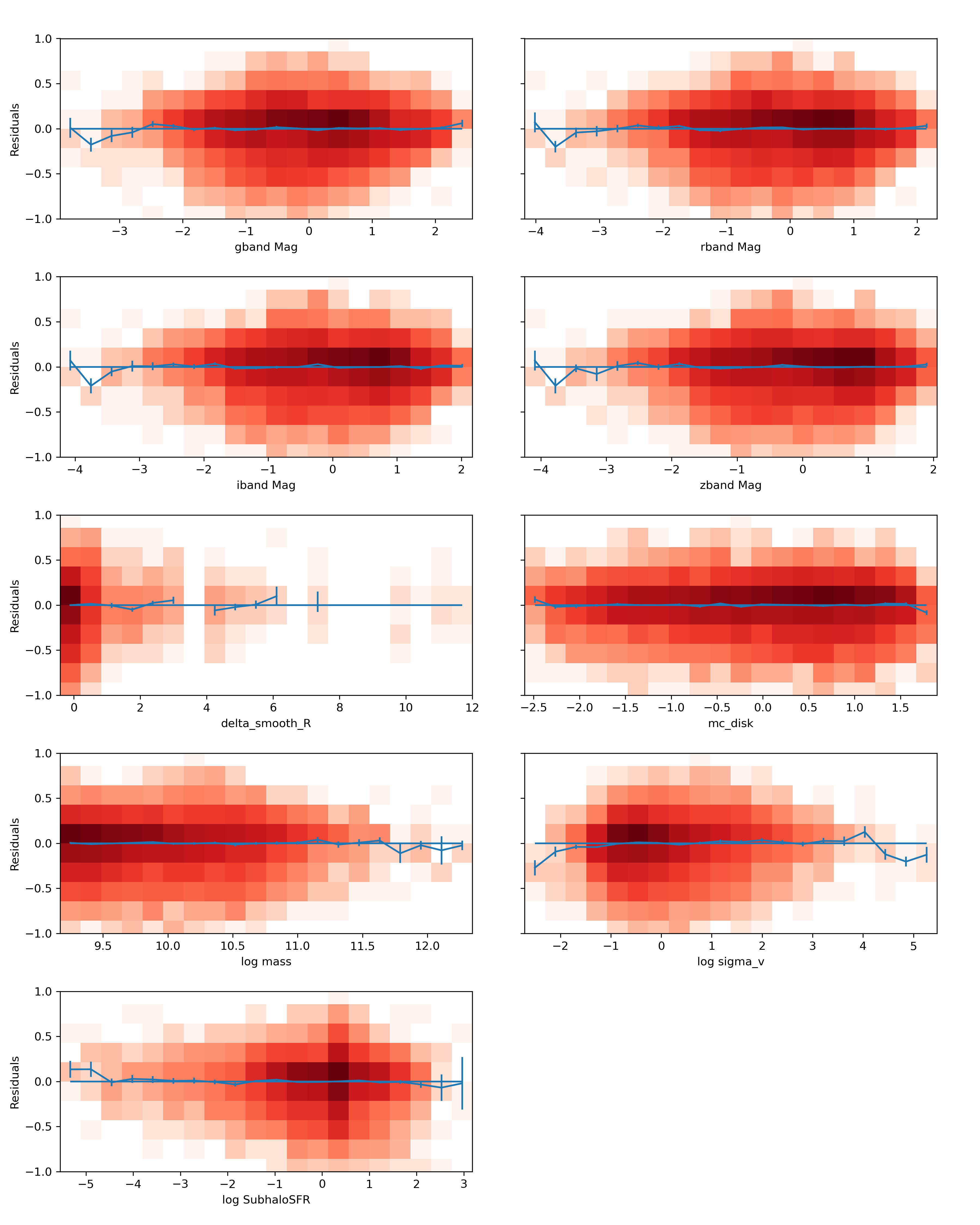}
  \caption{Residuals versus physical properties of the
  galaxies. The color scale reflects the density of residuals in that bin. The mean residual in a vertical slice is shown by the solid line, with error bars calculated by a jackknife procedure. The model has largely achieved its objective of removing any remaining correlation with these quantities.}
    \label{FullModelresidsvspreds}
\end{figure*}

Figure \ref{FinalModelTestSet}
depicts a fair amount of scatter around the
fitted line, but a central question is the following:
To what extent does this remaining scatter correlate
with physical properties of the galaxies, i.e.,
to what extent can the remaining scatter be
attributed to intrinsic properties of the galaxies?
Ideally, by incorporating these physical properties
into the model we have reduced any remaining such
correlation, and hence the residuals largely
originate from physical effects which occur between the galaxy and when it is observed.
Figure \ref{FullModelresidsvspreds} explores this by comparing the residuals
with galaxy features. It is observed that correlation in the residuals with each of the physical properties is largely eliminated.

\begin{figure*}
  \center
  \includegraphics[width=0.7\textwidth]{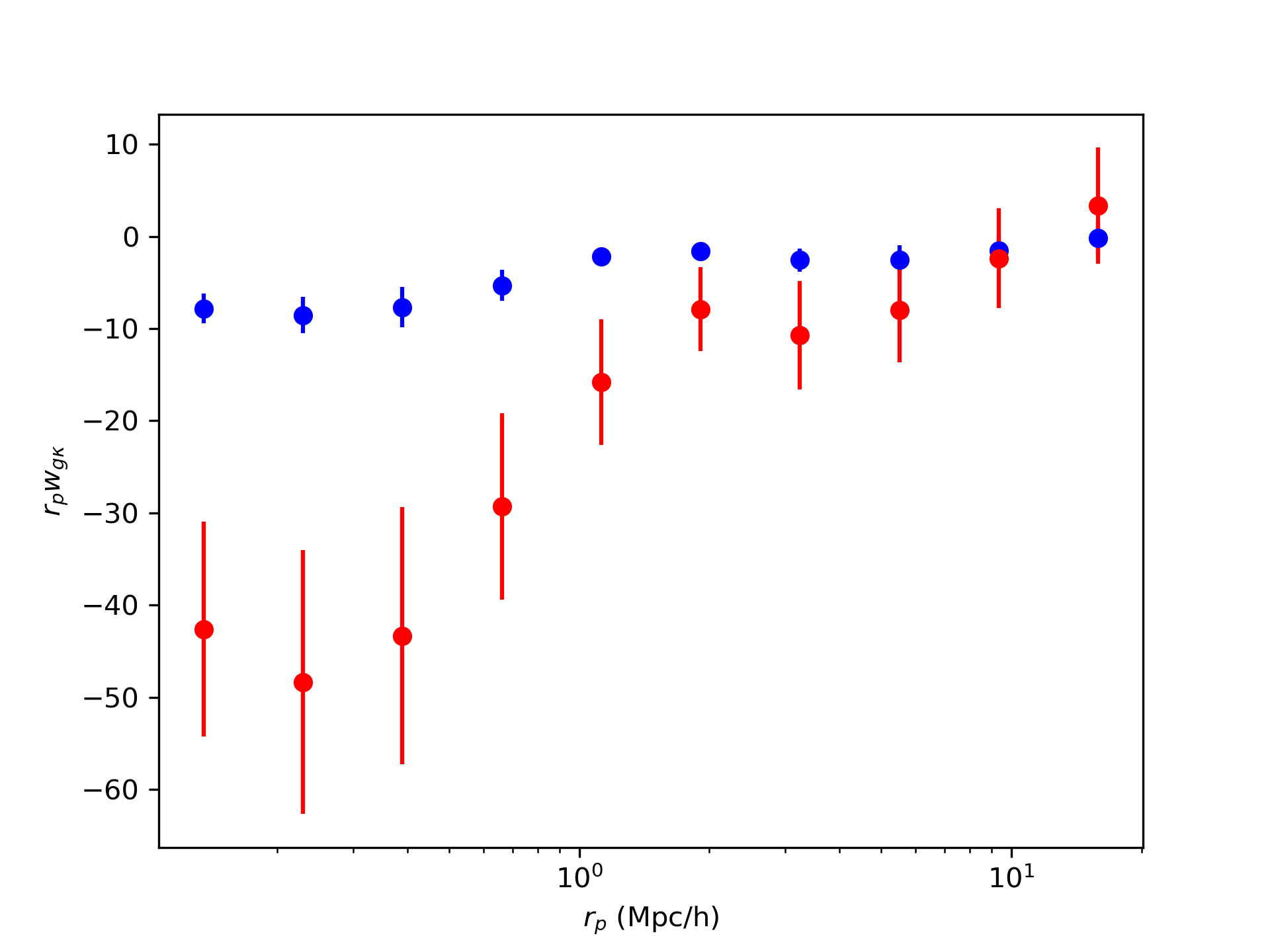}
  \caption{Spatial correlation in residuals from fit. A comparison
  is made between the baseline correlation in the data (in red),
  and the correlation in the residuals after the model is fit (in blue). While there is still
  remaining spatial correlation, especially on
  smaller angular scales, the degree of correlation
  has been reduced dramatically from that present
  in the original data.}
    \label{FullModelwgkappa}
\end{figure*}

Figure \ref{FullModelwgkappa} shows how the correlation
of residuals from the fit vary across scales. While
the reduction in the amount of spatial correlation
is encouraging, there remains a
clear, negative correlation on the smallest scales.
The pattern of correlation in this fit is consistent
with that seen in Figure \ref{residplot2}, indicating that
the additional complexity introduced in this final model
did not help to further reduce this correlation.
 This correlation at small scales is expected, since the galaxy physics at this scale is very difficult to capture. While our relatively simple model managed to reduce the correlation by a factor $\sim 5$, more sophisticated ML architectures may be needed in order to probe these small scale galactic physics, as demonstrated in \cite{graph-gan}, where adding graph convolutional layers to a neural network  removed remaining small scale correlations. Such approaches suffer, however, from reduced
 interpretability due to the convolutional abstraction of the inputs.
 

\section{Discussion}\label{discussion}

This work demonstrates the potential of supervised learning approaches that are designed to emphasize interpretability for yielding accurate predictions of intrinsic galaxy sizes. The residuals from such
a fit, estimates of the difference between the intrinsic and observed size, hold a wealth
of useful cosmological information
regarding topics such as
gravitational lensing and the peculiar velocity of galaxies. These techniques could  also potentially lead to improvements in
uncertainty quantification on photometrically-estimated redshifts.
The present work focuses on the use of data generated from the
simulation model Illustrus-TNG, in an effort to explore the limits of the potential
for such models, while also demonstrating a novel prediction
approach that incorporates the
learned structure in the high-resolution information
only available in the simulations.

The final model in this work serves as an illustration of the potential of
the developed methods, but also of the directions for further improvements.
The results demonstrate how photometry-only samples, in conjunction with
high-resolution simulation models, could be combined as part
of a framework to improve intrinsic galaxy size predictions. The model fits
yield a relatively interpretable picture of the way in photometrically-derived
properties relate to galaxy size. The results make it clear that magnitudes
are useful predictors of galaxy size, provided a sufficiently complex model
form is allowed. The residuals in the fits for intrinsic size show minimal
correlation with some key physical galaxy properties, indicating that the
models are successfully capturing the key relationships. It is clear, however,
that there remains correlation with local environment, and that photometric
data are not sufficient for capturing this correlation. This could possibly be
improved by using more complex supervised learning methods. The hope
is that the gain in interpretability from the proposed approach outweighs
the drawback of this remaining correlation.



A next step would be to explore the use of such
approaches with real, photometric survey data.
In such an analysis, the steps taken in this
work would be repeated, with Set 0 still
built from the simulation model. Set 1 should
consist of a sample of real galaxies with available
spectroscopic data, in order for reliable
measures of galaxy size and other observable galaxy properties to be available in the
training of the PPR model. This model could
then be applied to a photometry-only sample to
produce predictions of galaxy sizes. 
This approach would achieve the simultaneous goals
of using a modelling approach that produces
interpretable results, but also exploits the
information available in the high-resolution
simulation model. 



\section*{Data Availability Statement}

The data and software associated with this work are available on the World Wide Web. The simulation catalog data is available at \url{https://github.com/McWilliamsCenter/gal_decomp_paper}.
The software used for this work is available at \url{https://github.com/sukhdeep2/corr_pc}.

\section*{Acknowledgments}

CS is
supported by NSF Award Number 2020295.
SS is supported by McWilliams postdoctoral fellowship at CMU. YJ was supported in part by Department of Energy grant
DE-SC0010118 and in part by a grant from the Simons Foundation (Simons Investigator in Astrophysics, Award ID 620789). 

\bibliographystyle{mnras}
  \bibliography{refs,sukhdeep_FP}

\end{document}